\documentclass{PoS}

\usepackage{xspace} 
\newcommand{\Fermi}{\textit{Fermi}\xspace}

\newcommand{\adeg}{^\circ\!\!}
\newcommand{\bi}{\begin{itemize}}
\newcommand{\ei}{\end{itemize}}
\newcommand{\lb}{\label}

\title{Towards the First Catalog of \Fermi-LAT sources below 100 MeV}

\ShortTitle{Towards the First Catalog of \textit{Fermi}-LAT sources below 100 MeV}

\author{\speaker{Giacomo Principe}\thanks{On behalf of the Fermi-LAT collaboration.}\\
        Friedrich-Alexander-Universit{\"a}t Erlangen-N{\"u}rnberg, Erlangen Centre for Astroparticle Physics, \\
        Erwin-Rommel-Str. 1, 91058 Erlangen, Germany\\
        E-mail: \email{giacomo.principe@fau.de}}

\author{Dmitry Malyshev\\ %\thanks{A footnote may follow.}\\
        Friedrich-Alexander-Universit{\"a}t Erlangen-N{\"u}rnberg, Erlangen Centre for Astroparticle Physics, \\
        Erwin-Rommel-Str. 1, 91058 Erlangen, Germany\\
        E-mail: \email{dmitry.malyshev@fau.de}}

\author{Stefan Funk\\ %\thanks{A footnote may follow.}\\
        Friedrich-Alexander-Universit{\"a}t Erlangen-N{\"u}rnberg, Erlangen Centre for Astroparticle Physics, \\
        Erwin-Rommel-Str. 1, 91058 Erlangen, Germany\\
        E-mail: \email{s.funk@fau.de}}

\abstract{Previous analyses of point sources in the gamma-ray range were done either below 30 MeV or above 100 MeV. Below 30 MeV, the imaging Compton telescope (COMPTEL) onboard NASA's Compton Gamma-Ray Observatory detected 26 steady sources in the energy range from 0.75 to 30 MeV. At high energy, the \textit{Fermi} Large Area Telescope (LAT) has detected more than three thousand sources between 100 MeV and 300 GeV. 
Since the \textit{Fermi} LAT detects gamma rays also below 100 MeV, we apply a point source detection algorithm
in the energy range between 30 MeV and 100 MeV.
In the analysis we use PGWave, which is a background independent tool based on a wavelet transform. 
}

\FullConference{7th Fermi Symposium 2017\\
		15-20 October 2017\\
		Garmisch-Partenkirchen, Germany}

\begin{document}

\section{Introduction}
The Large Area Telescope \cite{Atwood:2000} 
on board the \textit{Fermi} gamma-ray space telescope has revolutionized our knowledge of the high energy sky.
The LAT detects $\gamma$-rays in the energy range from 20 MeV to more than 300 GeV, measuring their arrival times, energies, and directions.

% 3FGL catalog >100MeV
Although the LAT observations start at 20 MeV, all previous catalogs, released by the \textit{Fermi} Collaboration, are produced using optimized analysis focused on energies larger than 100 MeV.  
In particular, the Third \textit{Fermi}-LAT catalog \cite{Acero:2015hja}
characterizes 3033 sources in the energy range between 100 MeV and 300 GeV from the first 4 years of LAT data. 
Since the sensitivity of the instrument peaks at about 1 GeV, the 3FGL favours sources that are brightest around these energies.
At lower energies, COMPTEL analyzed the sources between 0.75 and 30 MeV \cite{Schoenfelder:2000bu}.

We are interested in studying the Fermi-LAT data between 30 MeV --100 MeV because they were not covered in the previous \textit{Fermi}-LAT catalogs. Since one of the largest uncertainties in the point source (PS) studies, below 1 GeV, is the uncertainty in the diffuse background, 
we use the PGWave tool, which is one of the background-independent methods, based on wavelet filtering to find significant clusters of gamma rays. 
We use PGWave to both detect the sources and to estimate their flux \cite{Principe:2017eox}.
%The first \textit{Fermi} Low Energy catalog (1FLE) is constructed using 8.7 years of LAT data taking full advantage of the improvements provided by the Pass 8 point spread function (PSF)-type event classification\footnote{A measure of the quality of the direction reconstruction is used to assign events to four quartiles}. 
%Special attention is given to the different PSF-type selections, in particular to the data cuts used to maximize the detection rate.
%In this proceeding we describe the method and performance of the analysis used to create the first catalog of \textit{Fermi}-LAT sources below 100 MeV. 
%We focus in particular on the develop of the point source analysis and we leave the presentation of the results for a following separate paper. 
%The first section contains an explanation of the point source analysis.
%The second section describe the simulation used for testing and optimizing the analysis.
%Finally the third section describe the efficiency in the point source detection and the consequently estimation of sensitivity of the catalog.

\section{Analysis procedure}
\label{analysis_procedure}
For the analysis, we use 12 regions of interest (ROIs) of size 180$^{\circ}$ in longitude and 90$^{\circ}$ in latitude. 
The centers of ROIs are at $b = 0^\circ,\; \pm 60^\circ$ and $\ell = 0^\circ,\; 90^\circ,\; 180^\circ,\; 270^\circ$.
The centers of the ROIs are chosen in order to cover the entire sky 
and to have an overlap between nearby regions of at least $30^{\circ}$. 

We represent the counts map using CAR projection \cite{Calabretta:2002vb} in Cartesian coordinates with pixel-size of $0^{\circ}\!\!.5$.
For each ROI, we perform the wavelet transform with the PGWave tool and eliminate the seeds that are closer to the border than 10$^{\circ}$ to avoid edge effects. 
We merge the seeds in the overlapping regions between different ROIs, eliminating duplicates within 2$^{\circ}$.

In order to avoid contamination due to the diffuse emission, we repeat the previous steps of the analysis also for the simulated maps of the Galactic and extra-galactic diffuse emission. 
In our list we eliminate the seeds that match those in the purely diffuse maps.
Thus, a diffuse emission model enters in the analysis indirectly: we eliminate point-like features in the diffuse emission 
from the list of the PS seeds.

Finally, we estimate the flux of the sources using the maxima in the wavelet transform (WT) map (referred to as WT peak values in the following for conciseness).
Diffuse emission can also affect the determination of the flux by introducing fluctuations of the wavelet transform map in addition to the Poisson noise.
We evaluate this effect using MC simulations by comparing the input and the reconstructed fluxes.
We also use MC simulations to find the optimal wavelet transform radii, which give a high detection efficiency while keeping the false positive rate at a few per cent level at high latitudes (see Section \ref{sec:PGWparam_selection}).
%For the search of PS in the data, we combine the results of wavelet transforms with two radii: $1^\circ\!\!.5$ and $2^\circ$. For radii smaller than $1^\circ\!\!.5$, the 68\% containment radius at 100 MeV is much larger than the wavelet transform scale, as a result, the efficiency of detection of faint sources becomes smaller. For radii larger than $2^\circ$, we start to lose sensitivity due to source confusion. In our analysis, the wavelet transform with the $2^\circ$ scale is important to detect faint sources at high latitudes, while the transform with the $1\adeg.5$ radius helps to improve source confusion, especially in the Galactic plane. In the analysis below, we combine the results of analysis with $1\adeg.5$ and $2^\circ$ wavelet scales.

\section{MC simulation and analysis optimization}
In this section we describe the simulation used for testing the analysis.
We explain the choice of the PSF event type, made in order to obtain the maximum detection rate for this energy range.
We use simulated gamma-ray maps to optimize the parameters in our PS detection algorithm.

\subsection{MC Simulation}
The large containment radius below 100 MeV implies that the number of independent positions in the sky, namely the positions that we could spatially distinguish with the given resolution, is not very large. 
Therefore, it is necessary to perform an accurate study of the event selection and of the parameters of the point-source analysis (PGWave parameter) in order to maximize the detection rate and minimize the number of false positives.
We use MC simulations to optimize the parameter of the analysis. 

The simulations have two steps: choice of the diffuse model and choice of the positions and fluxes of point sources.
For the diffuse model, we fit the gamma-ray data with a combination of templates that trace different components of emission,
such as hadronic interactions of cosmic ray with interstellar gas, inverse Compton (IC) scattering, \textit{Fermi} bubbles etc.
The construction of the diffuse model follows the same steps as the Sample model in \cite{TheFermi-LAT:2017vmf}.

We fit the data between 31.2 MeV and 312 MeV in 6 logarithmic bins.

\vspace{5mm}
\begin{table}[h]
\centering
\begin{tabular}{|c|c|}
\hline
Parameter & Value\\
\hline
\hline
Energy range & 30-100 MeV/ 100-300 MeV  \\
\hline
Pixel dimension & $0^{\circ}\!\!.5$\\
\hline
Interval of time & 8.7 years\\
\hline
Number of sources & 369\\
\hline
Flux & $10^{-8} - 10^{-4.5}{\rm \; cm^{-2}s^{-1}}$\\
\hline
\end{tabular}
\caption{ \label{simulation_map_par} Parameters used for the generation of MC maps.}
\end{table}

\begin{figure}[h]
\centering
\includegraphics[width=0.48\textwidth]{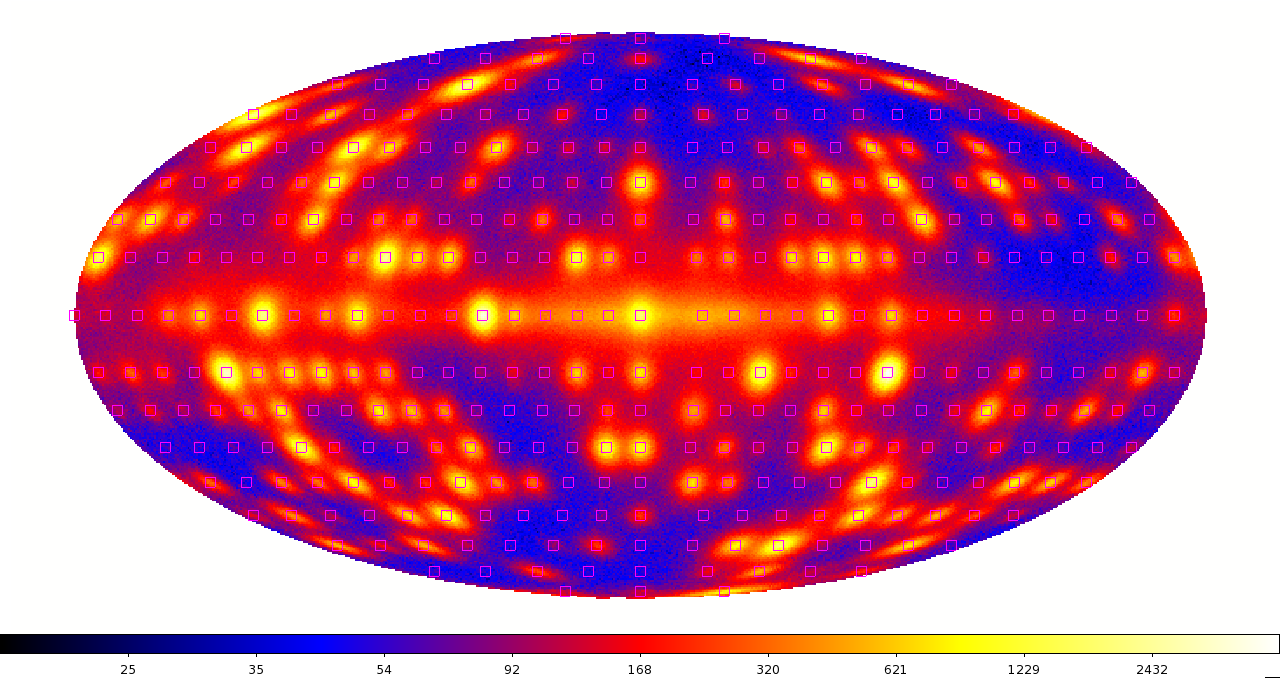}
\includegraphics[width=0.48\textwidth]{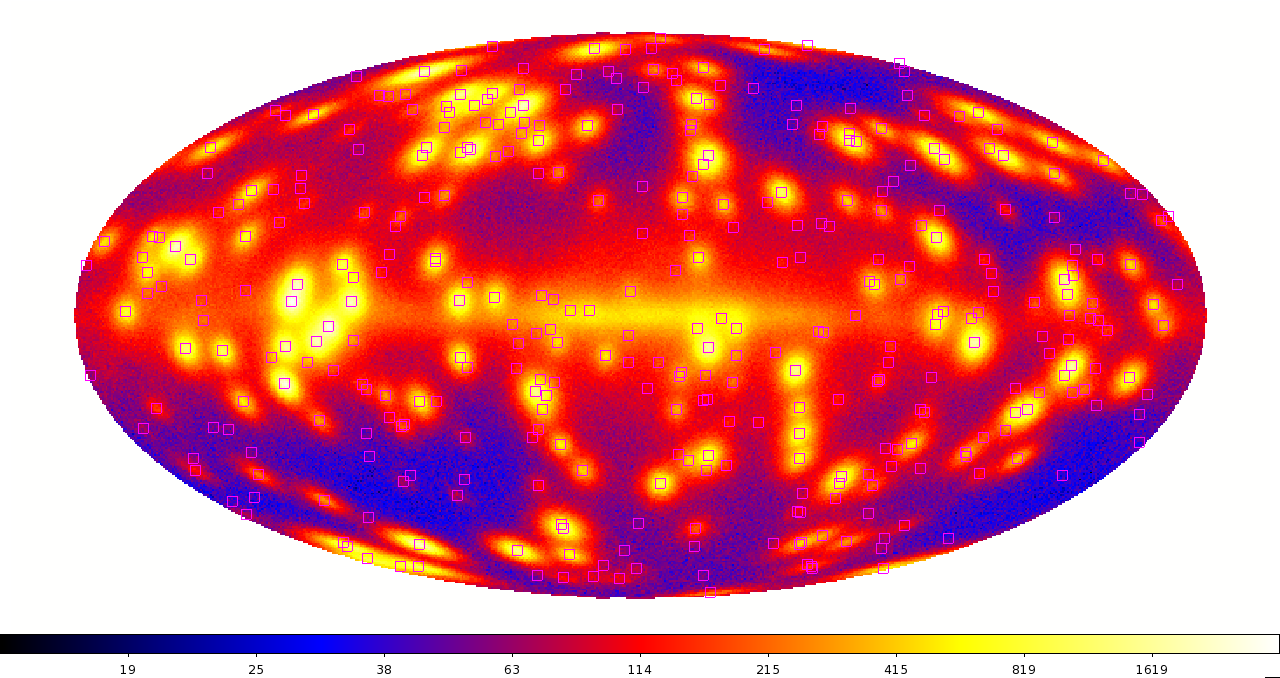}
\caption{\small \label{MC_map}
Left: counts map of the first setup (flat $S dN/dS$, positions on a grid)
with the simulated maps that contains 369 PS in a grid with 10$^{\circ}$ separation.
Right: counts map of the second setup (flat $S dN/dS$, random positions)
with the simulated maps that contains 369 PS randomly positioned in the sky. Both counts maps refer to the 30 -- 100 MeV band.}
\end{figure}

For the generation of point sources, we use two different setups. 
Both setups contain a population of 369 PS with a flux between 30 and 100 MeV randomly chosen from 
a flat distribution on the logarithmic scale between 
$10^{-8}{\rm \; cm^{-2}s^{-1}}$ (which is close to the threshold of the detection) and $10^{-4.5}{\rm \; cm^{-2}s^{-1}}$.
In the first setup the PS are positioned in the sky in a grid with a separation of $10^{\circ}$ (Figure \ref{MC_map}, left), while in the second one they are randomly positioned in the sky (Figure \ref{MC_map}, right).
Figure \ref{MC_map} and all the following figures refer to the energy bin 30 -- 100 MeV. 
Table \ref{simulation_map_par} contains the main parameters that are used in the simulation.

\subsection{Selection of event type}
\label{sec:PSF_selection}

%Here we describe the Selection of the PSF classes.
At low energies, \textit{Fermi} LAT has a PSF that increases from $\sim 3^\circ$ at 100 MeV to $\sim 10^\circ$ at 30 MeV.
One can improve the resolution by selecting events with a better angular resolution (e.g., PSF3 event type),
but this subselection of events leads to a decrease of statistics.

In order to find the optimal combination of PSF event types, 
we compare the detection efficiencies and false positive rates for the following selections:
\bi
\item
all PSF event types combined together;
\item
PSF1, PSF2 and PSF3 event types; 
\item
PSF2 and PSF3 event types;
\item
only PSF3 event type.
\ei
We use the same analysis pipeline that is described in Section \ref{analysis_procedure}. 
One of the most important parameters is the wavelet transform scale.

\begin{figure}[h]
\centering
\includegraphics[width=0.49\textwidth]{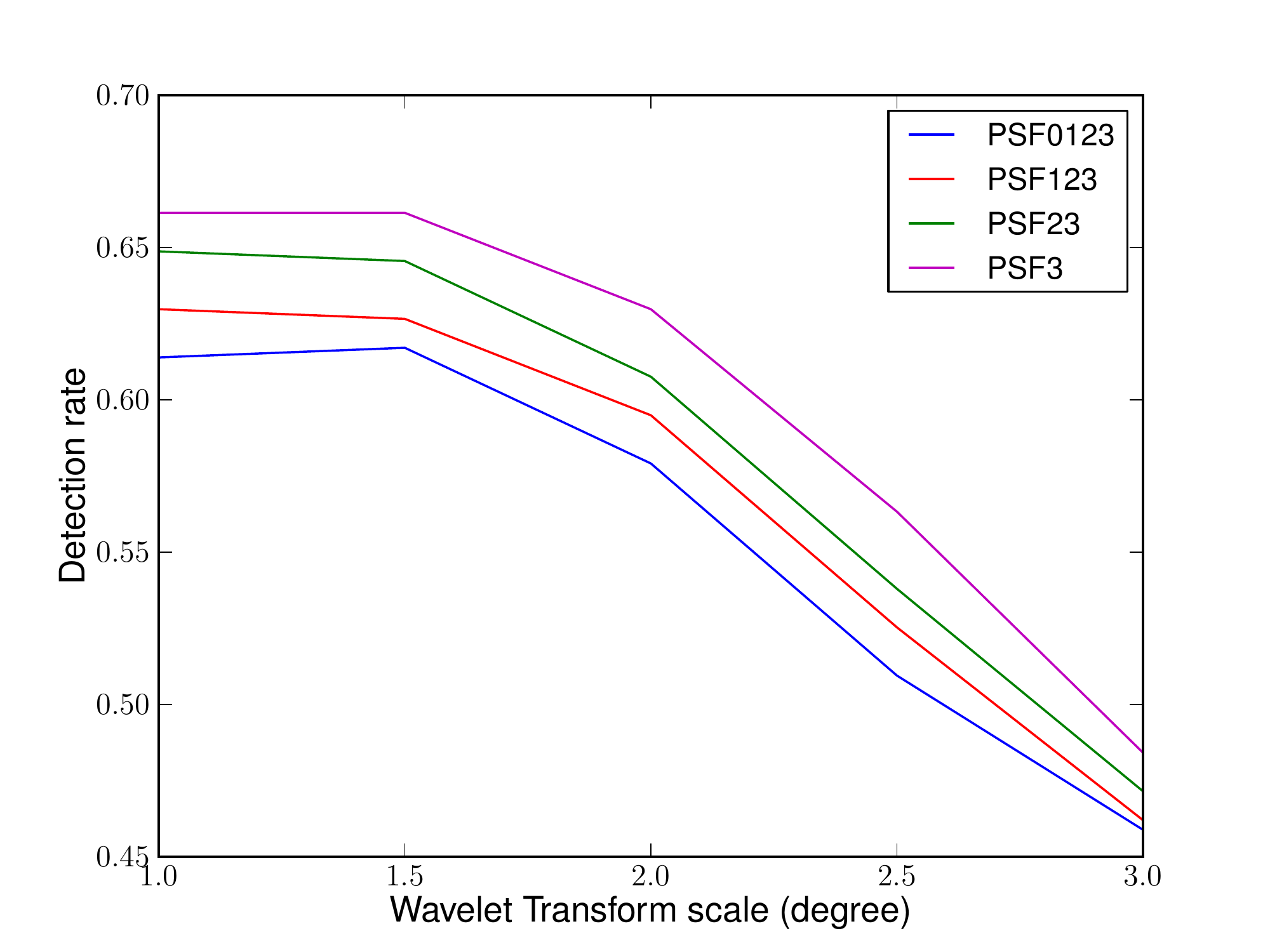}
\includegraphics[width=0.49\textwidth]{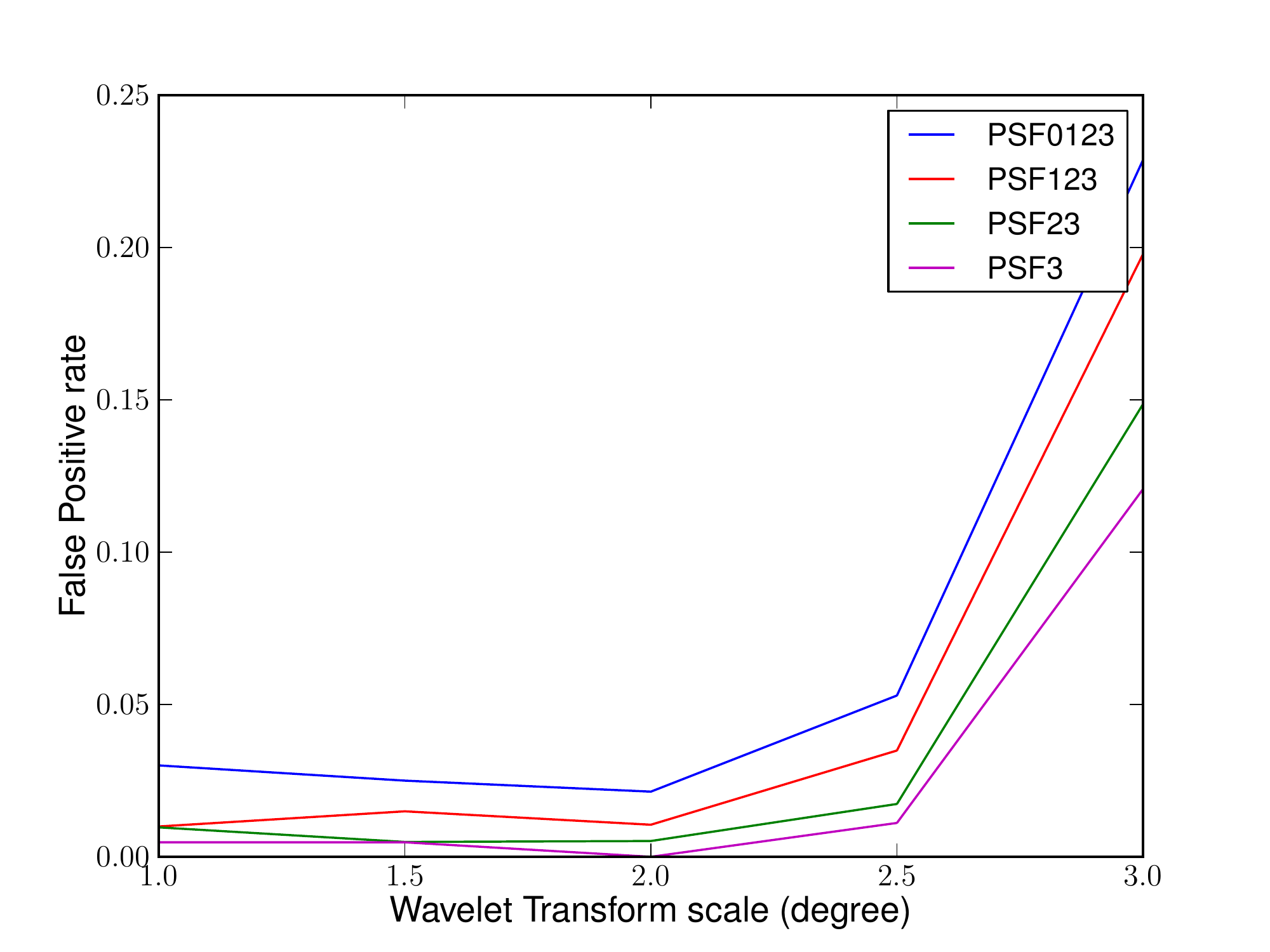}
\caption{\small \label{grid_detection}
Left: 
detection rate varying the dimension of the wavelet transform scale (flat $S dN/dS$, random positions).
Right: 
false positive rate varying the dimension of the wavelet transform scale (flat $S dN/dS$, random positions).
Both plots refer to the 30 -- 100 MeV band.
}
\end{figure}

Figure \ref{grid_detection} shows the detection rate and the false positive rate, using the grid setup, for different combinations of PSF event types, where we vary the wavelet transform scale. 
%They show respectively the detection rate and the false positive rate as a function of the PGWave radius.
The plots show that for the PSF3 event type the detection rate is the largest while the false positive rate is the smallest. A similar behavior is observed in the analysis using the random setup. 
Both the grid and the random positions setups show that the PSF3 event type, even if it has smaller statistic, gives the best detection and false positive rates.

\subsection{Selection of PGWave parameters}
\label{sec:PGWparam_selection}

In order to maximize the detection rate and minimize the false positive rate, we optimize the main parameters of our analysis.
For this analysis we use the simulated maps with sources with a flux extrapolated from the 3FGL, and randomly positioned in the sky.
The parameters that we optimize are reported in Table \ref{pgw_par}. The threshold was set at 3 $\sigma$. 
We perform the analysis separately for both energy bins: 30 -- 100 MeV and 100 -- 300 MeV.

We study the behavior of detection rate and false positive rate varying the wavelet transform scale, 
the minimum number of connected pixel to define a peak and the minimum distance between the sources. 
We vary also the merging radius, that is the tolerance radius for merging the seeds from different ROIs (see analysis description).

\vspace{5mm}
\begin{table}[h]
\centering
\begin{tabular}{|c|c|c|}
\hline
PGWave parameter & Values & Step\\
\hline
\hline
MH Wavelet Transform scale & $1^{\circ} - 3\adeg.5$ & $0\adeg.5$  \\
\hline
Min. number of connected pixels & $2 - 8$ & 1 \\
\hline
Min. distance between sources & $2^{\circ} - 3^{\circ}$ & $0\adeg.5$\\
\hline
\end{tabular}
\caption{ \label{pgw_par} List of PGWave parameters used to optimize the analysis. 
%Considering also the variation of the merging radius, we try more than 170 different combinations of the analysis parameters.
}
\end{table}

\begin{figure}[h]
\centering
\includegraphics[width=0.49\textwidth]{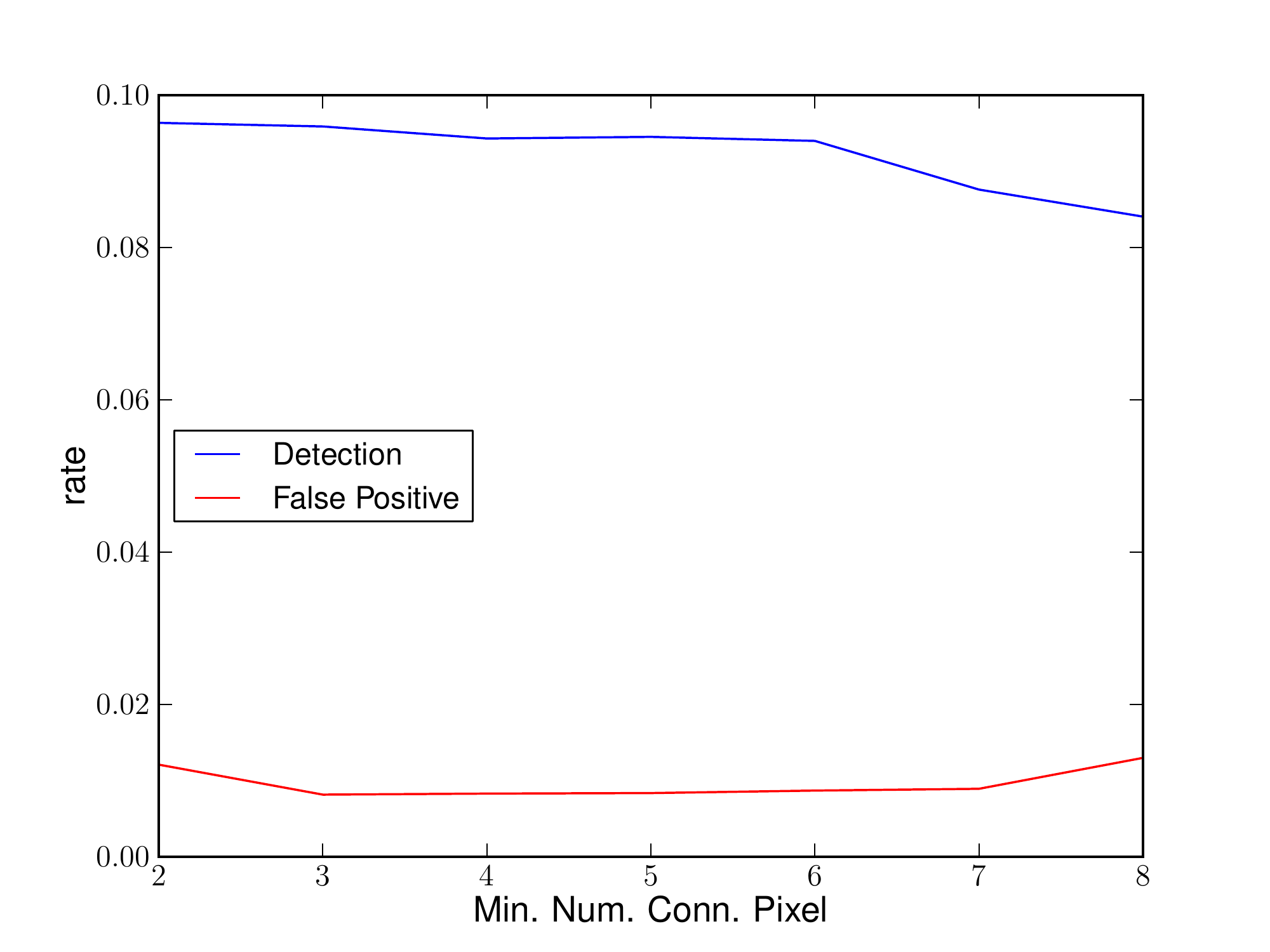}
\includegraphics[width=0.49\textwidth]{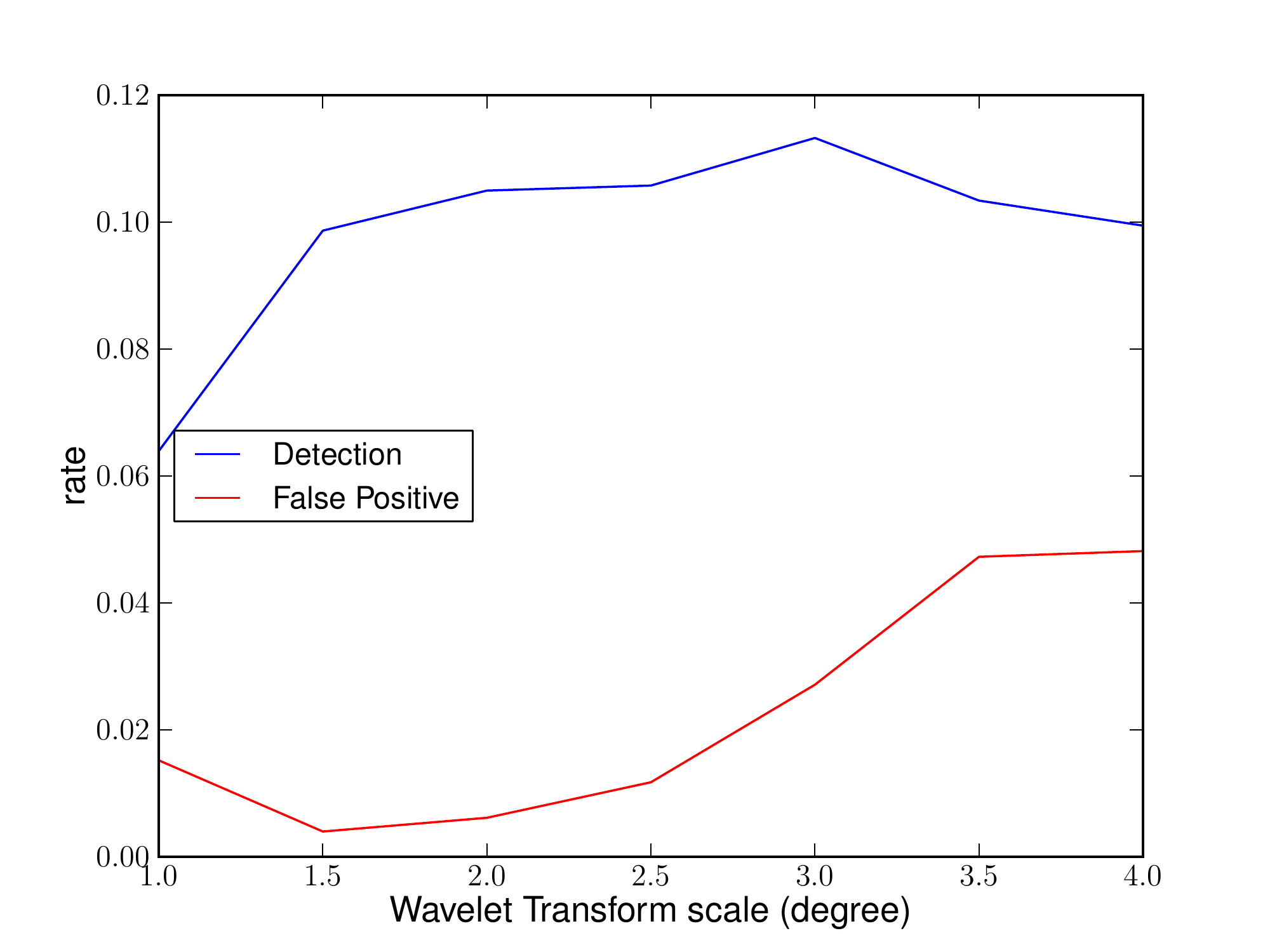}
\caption{\small \label{optimize_ex}
Left: 
detection and false positive rates as a function of the minimum number of connected pixels (extrapolated 3FGL sources with randomized position).
Right: 
detection and false positive rates as a function of the PGWave WT scale (extrapolated 3FGL sources with randomized position). Both plots refer to the 30 -- 100 MeV band.
}
\end{figure}

Figure \ref{optimize_ex} on the left shows the dependence of the detection and false positive rates on the minimum number of connected pixels,
while Figure \ref{optimize_ex} on the right shows the rates as a function of the wavelet transform scale.

We chose to use in the analysis of the data the PGWave parameters that maximize the detection rate and minimize the false positive rate.
The optimal values of the parameters are reported in Table \ref{pgw_final_par}.
For the analysis of the data, we chose to use two different wavelet scales: $1\adeg.5$ (important to resolve source confusion, especially in the Galactic plane) 
and  $2^\circ$ (important to detect faint sources at high latitudes).
They give the false positive rate for high latitude sources, namely with $\mid b \mid > 10 ^\circ$, lower than 2\% (6\%) in the energy bin 30 -- 100 MeV (100 -- 300 MeV).

\begin{table}[h]
\centering
\begin{tabular}{|c|c|}
\hline
PGWave parameter & Chosen value\\
\hline
\hline
MH Wavelet Transform scale &  $2\adeg.0$ ($1\adeg.5$) \\
\hline
Minimum number of connected pixels & 6 (5)\\
\hline
Minimum distance between different sources & $3\adeg.0$ ($2\adeg.5$)\\
\hline
\end{tabular}
\caption{\label{pgw_final_par} List of PGWave parameter resulting from the optimization. We use these values for the analysis of the data. In parenthesis the PGWave values of the second choices for the combined analysis.}
\end{table}

\section{Detection Efficiency}
\lb{sec:det_eff}
For the determination of the detection efficiency, we used MC maps that include diffuse emission and PS with random positions in the sky.
The spectra of the input PS are determined from extrapolation of the spectra of PS randomly selected from the 3FGL catalog.

We apply the analysis to the simulated maps and we compare the resulting sources with the list of input sources in order to estimate the detection efficiency and the false positive rate, 
which are the ratio between the number of detected sources and the input sources and the ratio between the PGWave seeds that do not have an association and the total number of seeds.

\begin{figure}[h]
\centering
\includegraphics[width=0.49\textwidth]{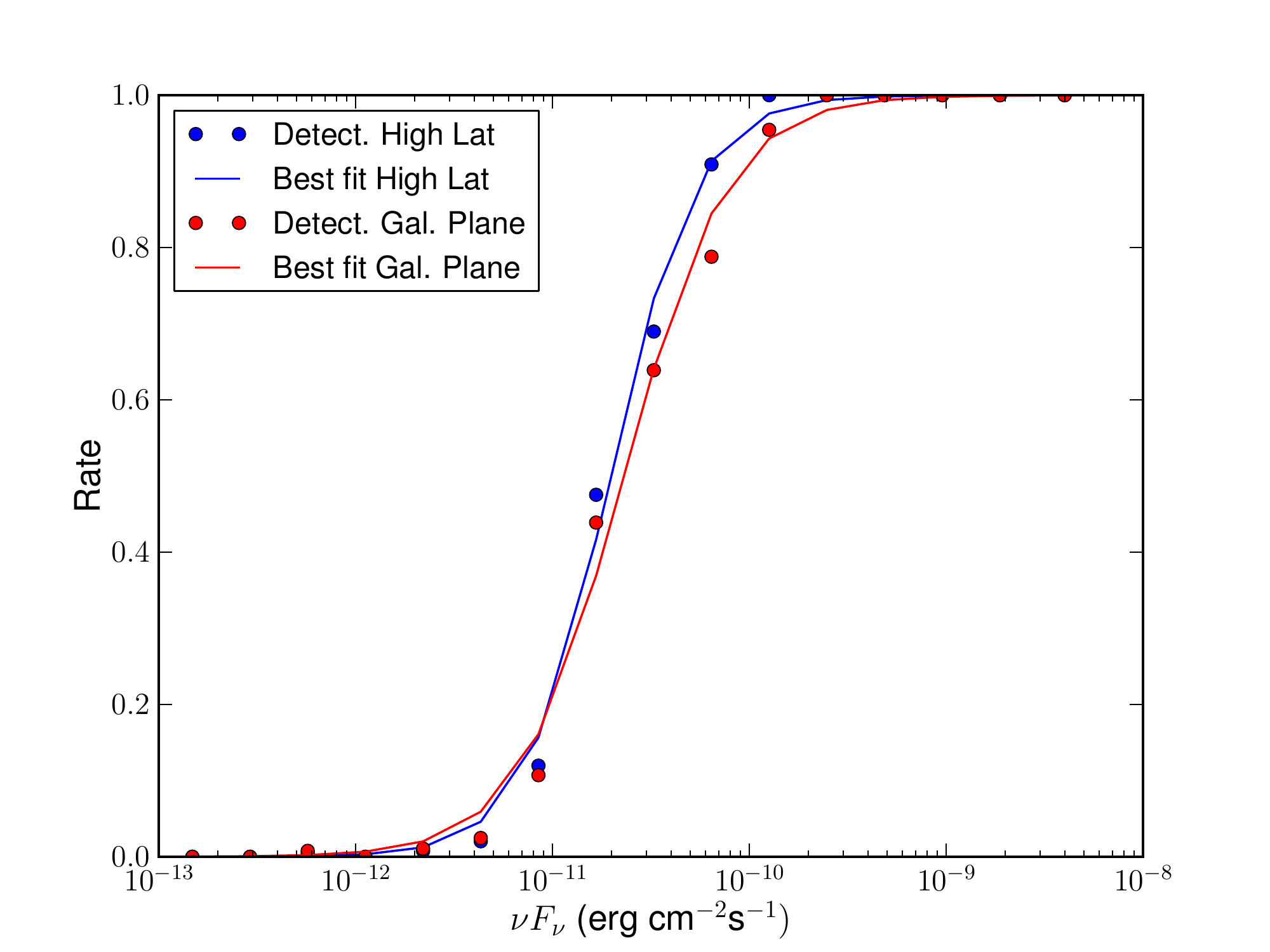}
\includegraphics[width=0.49\textwidth]{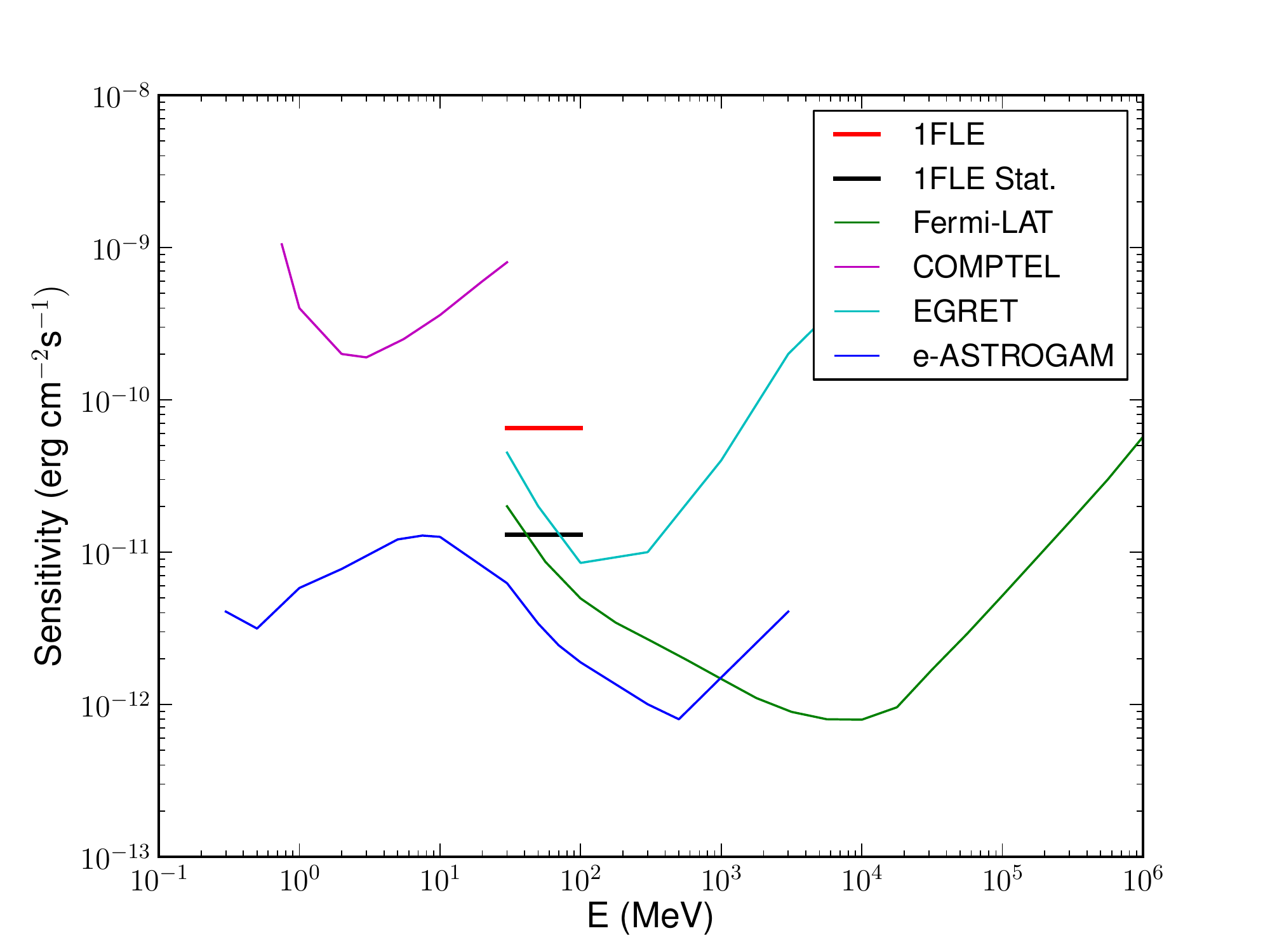}
\caption{\small \label{detection_flux}
Left: 
detection rate as a function of the input flux of the simulated PS with random position in the sky for high latitude sources, $\mid b \mid > 10 ^\circ$. 
Right: 
Comparison of the PS sensitivity of the 1FLE catalog and the differential sensitivities of different $\gamma$-ray instruments. 
The COMPTEL (magenta line) and EGRET (cyan line) sensitivities are given for the typical observation time accumulated during the $\sim$ 9 years of the CGRO mission. The \textit{Fermi}-LAT sensitivity (green line) is for a high Galactic latitude source in 10 years of observation in survey mode. The blue line represents the e-ASTROGAM sensitivity for an effective exposure of 1 year and for a source at high Galactic latitude \cite{DeAngelis:2016slk}. 
In red, the 1FLE total sensitivity determined from the 95\% detection rate in the plot on the left, 
while the black represents the 1FLE statistical sensitivity determined as the flux corresponding to the 5$\sigma$ significance of PGWave.
}
\end{figure}

%Results comparison with Input MC
Figure \ref{detection_flux} on the left shows the ratio of detected sources to the input MC sources in each of the flux bins for high latitude sources, namely $\mid b \mid > 10 ^\circ$.
%The plots show that PGWave is sensitive only to sources with a flux above $\sim 10^{-6}{\rm \; cm^{-2}s^{-1}}$.
We model the detection efficiency by a hyperbolic tangent function 
$({1 + \tanh\lambda(f - f_0)})/{2},$
where parameters $\lambda$ and $f_0$ are 
determined by fitting to the detection efficiency points.
Using this model, 
we find that PGWave has a detection rate larger than 95\% for PS with a energy flux higher than $6.5 \times 10^{-11}{\rm \; erg \; cm^{-2}s^{-1}}$ at 55 MeV.

\section{Conclusion: 1FLE sensitivity}
%The aim of this catalog is to close the gap between the previous $\gamma$-ray catalogs:  the COMPTEL observations at energies lower than 30 MeV and the \textit{Fermi}-LAT catalogs at energies higher than 100 MeV.

% refer to eASTROGAM
%The 1FLE catalog confirms the importance of having an instrument with a better angular resolution at energies below 100 MeV in order to increase the number of sources detected at MeV energies, such as e-ASTROGAM \cite{DeAngelis:2016slk} or AMEGO\footnote{See \small https://pcos.gsfc.nasa.gov/physpag/probe/AMEGO\_probe.pdf}.
In Figure \ref{detection_flux} right, we compare the sensitivity of the 1FLE catalog to the PS sensitivities of various gamma-ray experiments.
%Notice, that the 1FLE sensitivity includes systematic uncertainty.
The red line is the full 1FLE sensitivity calculated from 95\% detection efficiency in MC tests at latitudes $|b| > 10^\circ$
(Section \ref{sec:det_eff}).
The black line is an average flux corresponding to $5\sigma$ statistical significance of the wavelet transform peak value. 
The full sensitivity is about 5 times worse than the statistical one due to additional selection criteria, 
such as the minimal number of connected pixels in PGWave, 
which are used to reduce the false positive rate.
%At energies below 100 MeV, both e-ASTROGAM and AMEGO are expected to have a sensitivity which is more than two times better than that of \textit{Fermi}-LAT. These new instruments are able to widely extend our knowledge of the MeV sky.
We find that PGWave provides a reasonable detection efficiency and low false positive rate, 
which make it a useful tool for an analysis of \Fermi LAT PS below 100 MeV in a background independent way.

\section*{Aknowledgements}
The \textit{Fermi}-LAT Collaboration acknowledges support from NASA and DOE (United States), CEA/Irfu, IN2P3/CNRS, and CNES (France), ASI, INFN, and INAF (Italy), MEXT, KEK, and JAXA (Japan), and the K.A. Wallenberg Foundation, the Swedish Research Council, and the National Space Board (Sweden).

\end{document}